# A 140-GHz Direct Raised-Cosine Envelope-Shaping Transmitter with Integrated ILO Phase Shifter

Haoling Li, Najme Ebrahimi
Northeastern University, USA
{li.haoling, n.ebrahimi}@northeastern.edu

***Abstract*—A 140-GHz transmitter with direct raised-cosine-like envelope shaping and wide-range phase tuning is presented in 90-nm SiGe BiCMOS. The proposed architecture relaxes conventional baseband pulse shaping and high-speed digital-to-analog converters by directly synthesizing 3-level and 5-level RF envelope states that approximate a raised-cosine waveform, enabled by a 23-dB modulation dynamic range. Measured results demonstrate data rate up to 8 Gb/s with 34-dB sidelobe suppression. To support scalable phased-array applications, an injection-locked-oscillator phase-tuning path is also integrated, providing 22.5° digital phase resolution together with continuous analog tuning over more than ±360°. The transmitter achieves up to 2 dBm output power and combines direct RF envelope shaping with wide-range fine phase control for sub-THz wireless links.**

## I. Introduction

The growing demand for multi-gigabit wireless links has motivated increasing interest in the sub-terahertz (sub-THz) spectrum. In particular, the D-band (110–170 GHz) offers large available bandwidth for short-range high-capacity links, including point-to-point backhaul, local-area wireless connectivity, and intra-data-center communication [1-8]. Direct modulation using on-off keying (OOK) can improve transmitter efficiency by avoiding complex digital baseband processing. However, without pulse shaping, OOK produces a sinc-like spectrum with sidelobes typically only 13 dB below the main lobe, which increases out-of-band (OOB) emissions and causes adjacent-channel leakage [1, 2], Fig. 1(a). This spectral leakage poses a growing problem in multi-link deployment scenarios. In dense urban small-cell backhaul, unshaped sidelobes from each transmitter degrade the signal-to-interference ratio of neighboring channels. A similar constraint arises in hyperscale data centers, where sub-THz wireless flyways between top-of-rack switches share a confined, multipath-rich environment [3], and each additional unshaped link raises the aggregate interference floor for all co-located channels. In both applications, the number of simultaneously usable frequency slots, and hence the multi-channel spatial reuse, is limited not by available bandwidth but by OOB sidelobe levels.

Previous works have demonstrated direct OOK modulation using amplifier-based switching [1, 2], Fig. 1(a), including VCOs with embedded transmission-line modulation [4], and harmonic- or multiplier-based on-off keying [5-7], achieving data rates up to 20 Gb/s. Since these approaches rely on abrupt switching without pulse shaping, their sidelobe suppression typically remains around 15 dB, which limits the number of usable adjacent channels in multi-link deployments. More recently, [9] demonstrated root raised cosine (RRC) pulse-shaped QPSK at 25 GHz using digitally controlled phase shifters and attenuators in a commercial off-the-shelf beamformer IC without high-speed Digital-to-Analog Converters (DACs). At baseband, [10] implemented raised-cosine (RC) synthesis using digital up sampling and a non-binary DAC, improving spectral roll-off of 36 dB at 24 GS/s at the cost of added circuit complexity and power.

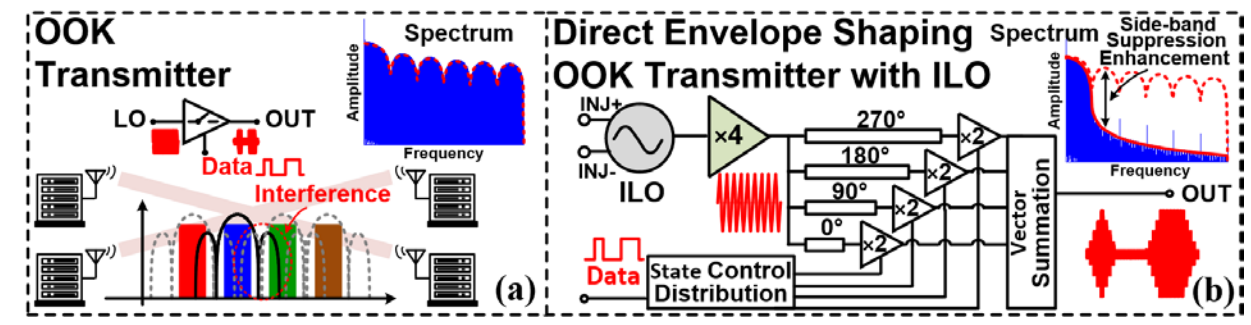

Fig. 1: Conventional direct-switching OOK transmitter. (b) Proposed RC-OOK transmitter with direct raised-cosine envelope shaping via quadrature reconfigurable OOK-modulator paths.

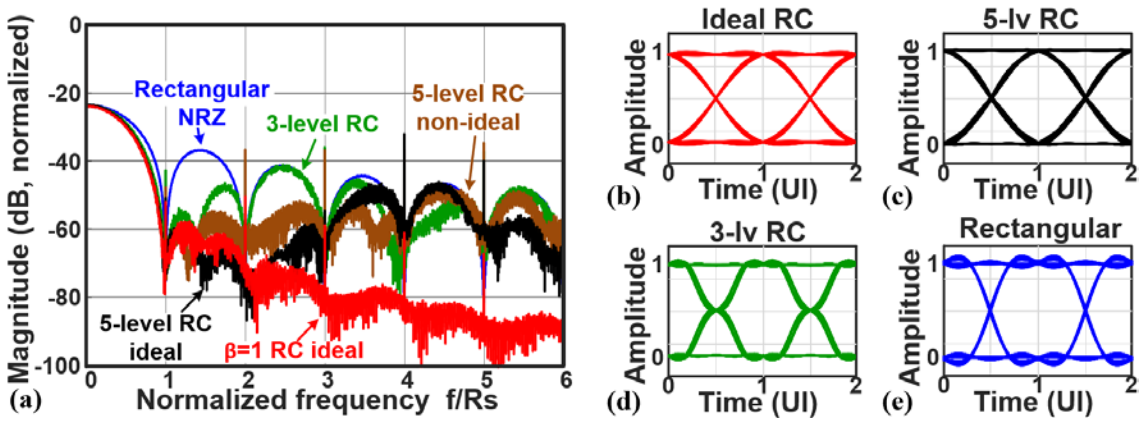

Fig. 2: Simulated comparison of spectral and time-domain performance. (a) Output spectra for rectangular NRZ, ideal RC, 3-level RC, and ideal/nonideal 5-level RC waveforms. (b)–(e) Corresponding eye diagrams under a receiver bandwidth of 2.5× the symbol rate, showing progressively improved eye quality and reduced ISI from rectangular NRZ to 5-level RC-OOK.

In this work, as illustrated in Fig. 1(b), the proposed transmitter performs direct RC envelope shaping at 140 GHz by synthesizing multi-level amplitude states within each OOK symbol through selective activation of quadrature harmonic-modulator paths. The effect of different RC approximations on spectral sidelobes and eye diagrams is shown in Fig. 2, demonstrating >30-dB sidelobe suppression and reduced ISI/timing jitter compared with the rectangular waveform. The required multi-level modulation dynamic range is enabled by the Double-HOOK [5] structure with embedded phase shifting and vector summation. To support scalable phased-array beamforming and carrier spectral purity, a D-band injection-locked oscillator (ILO) phase shifter is integrated, providing 22.5° digital phase resolution and continuous analog varactor tuning beyond ±360° through the ×8 multiplication chain. Section II describes the proposed direct RC transmitter architecture, Sec. III discusses circuit implementation, Sec. IV presents measurement results, and Sec. V concludes the paper.

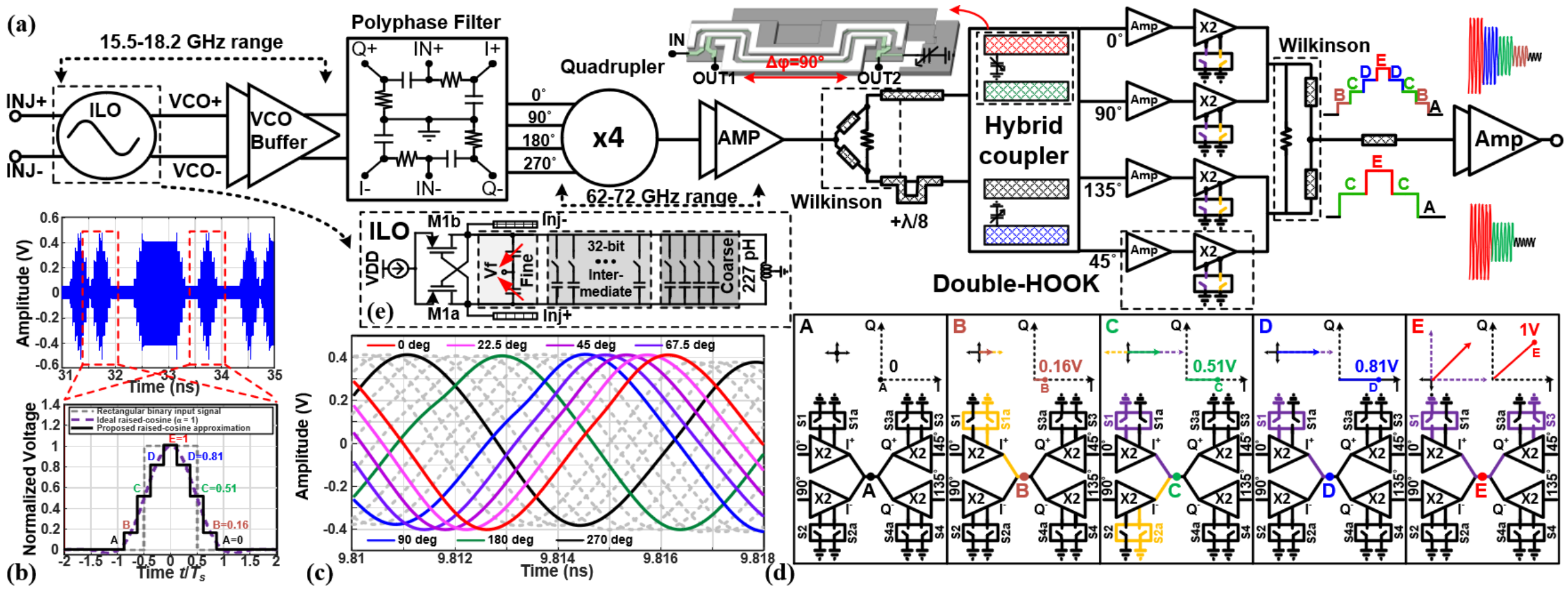

Fig. 3: Proposed RC-OOK transmitter, (a) block diagram of transmitter with ILO, ×4 multiplier, quadrature generation, and staircase path selection modulator. (b) Ideal RC envelope versus 5-level staircase approximation. (c) Simulated ILO phase tuning showing 360° range. (d) Switch-state combinations producing five amplitude levels over 25.5-dB dynamic range. (e) ILO simplified core schematic with capacitor-bank and varactor tuning.

## II. Proposed Direct Envelope-Shaping Modulator

### A. *Direct Raised Cosine Envelope Shaping*

The RC response is a standard Nyquist pulse-shaping function that provides a band-limited spectrum with smooth inter-symbol transitions. Its time-domain impulse response is given by Eq. (1),

$$g_{rc}(t) = \frac{\sin\left(\frac{\pi t}{T_s}\right)}{\frac{\pi t}{T_s}} \cdot \frac{\cos\left(\frac{\pi \alpha t}{T_s}\right)}{1-\left(\frac{2\alpha t}{T_s}\right)^2} \tag{1}$$

where $T_s$ denotes the symbol period and $\alpha$ is the roll-off factor controlling the excess bandwidth ($0 \le \alpha \le 1$). In this implementation, $\alpha = 1$ is selected to maximize the roll-off bandwidth, which relaxes the timing precision required for each staircase step. With $\alpha = 1$, the main lobe occupies twice the minimum Nyquist bandwidth, but the sidelobe decay rate is maximized. The ideal RC envelope with $\alpha = 1$ suppresses OOB sidelobes much better than the rectangular sinc case, whose first sidelobe is 13 dB down. When sampled at $T_s/4$ intervals, the ideal normalized RC envelope produces five amplitude levels: 0, 0.170, 0.500, 0.849, and 1.000. The transmitter synthesizes levels of 0, 0.160, 0.510, 0.810, and 1.000, deviations of 5.9%, 2.0%, 4.6%, and 0%, respectively. As shown in Fig. 3(b), the resulting 5-level staircase closely follows the ideal RC envelope. Fig. 2 illustrates the spectral and time-domain consequences of this approximation. The rectangular waveform exhibits the strongest OOB emissions due to its slow 1/f sidelobe decay. The proposed 5-level staircase approaches the ideal RC spectrum, while the 3-level approximation and non-ideal 5-level cases show progressively higher sidelobes. The eye diagrams in Fig. 2(b)–(e), obtained under a receiver bandwidth of 2.5× the symbol rate, confirm that more accurate multi-level approximations yield cleaner eye diagrams with reduced ISI and timing jitter.

### B. *Multi-amplitude-level synthesis for RC Envelope shaping*

The core of the proposed RC-OOK transmitter is a quadrature second-harmonic modulator with 25.5-dB modulation dynamic range. A passive quadrature network splits the LO into four branches at 0°, 45°, 90°, and 135°, each driving a push-push doubler with two embedded emitter-shunt switching devices of different sizes (10 µm and 0.6 µm), [6]. The larger device provides the primary ON/OFF modulation depth, while the smaller device introduces a fine amplitude offset, yielding three distinct impedance states per branch. By selectively activating combinations of these states across the four quadrature paths and exploiting vector summation at the output combiner, the transmitter synthesizes five amplitude levels spanning the full 25.5-dB range, as illustrated in Fig. 3(b) and (d). Level A (−22 dBm) corresponds to all paths disabled; Level B (−10.5 dBm) activates only the fine-control device in one path; Level C (−2 dBm) combines the primary and fine-control devices in complementary paths; Level D (1.8 dBm) activates one primary device alone; and Level E (+3.5 dBm) engages primary devices in both I and Q branches for coherent quadrature combining at full power. For three-level RC-OOK at higher data rates, only levels A, C, and E are used. Multiple equivalent switch-state combinations exist for each target level, providing flexibility for calibration and mismatch compensation.

## III. Circuit Implementation of Proposed Transmitter

The proposed RC-OOK transmitter, shown in Fig. 3(a), is composed of following blocks: an ILO based on [11] for carrier generation with phase shifting and improved carrier phase purity, multiplier by 4 (×4) LO generation, a quadrature second-harmonic (×2) modulator for multi-level RC envelope synthesis, and output combining with amplitude calibration. The ILO core, shown in Fig. 3(e), uses a 227-pH inductor with three tuning mechanisms: a 4-bit MIM capacitor bank for coarse frequency selection, a 32-bit capacitor bank for intermediate tuning, and an analog nMOS varactor for continuous fine adjustment. Under $f_{inj} = 19.3\ GHz$ and an injection power of $P_{inj} = -10\ dBm$, the simulated tuning range spans 16.5–19.8 GHz with up to 50° phase shift at the fundamental. As shown in Fig. 3(c), digital and analog tuning combined provide a continuous phase range exceeding ±360° through the ×8 chain, with a simulated locking range of approximately 300 MHz. The ILO output is buffered and split into I and Q components by a two-stage polyphase filter achieving less than 2° phase imbalance and under 1 dB amplitude mismatch across 16.5–19.8 GHz. The filter introduces 9 dB loss, recovered by 19 dB of buffer gain for 10 dB net gain. A four-path ×4 multiplier with 1 dB loss, followed by wideband buffers, generates the 62–72 GHz LO delivering up to 10.2 dBm while consuming 32 mW DC.

Accurate quadrature LO generation is essential for the multi-level vector summation that produces the RC envelope. The 62–72 GHz LO drives the RC-OOK modulator through a phase-generation network. The Wilkinson divider with a λ/8 delay line feeds a pair of 90° hybrid couplers, producing four excitation phases at 0°, 45°, 90°, and 135°. A buffer amplifier (8 dB gain, 0.9 V bias, 1.6 V supply) compensates the network loss. Each branch drives a differential balun and a push-push SiGe HBT doubler for second-harmonic generation at 124–144 GHz. The doublers use 6-µm HBTs biased in Class-B (0.7 V base, 1.6 V supply) to maximize harmonic conversion while suppressing fundamental leakage that would corrupt the multi-level amplitude accuracy. The 10-µm and 0.6-µm emitter-shunt devices, shunted by stub transmission lines to suppress off-state parasitics, provide the three impedance states per path enabling the 5-level RC envelope shaping described in Section II-B. The I and Q pairs are combined by a Wilkinson combiner, followed by a two-stage buffer providing 8 dB gain at 12 mW dissipation and boosting maximum output to 3.5 dBm. The simulated output power remains flat across the LO tuning range with less than 3 dB variation, ensuring consistent envelope accuracy. To maintain the amplitude precision required for strong sidelobe suppression, two calibration mechanisms are integrated: an nMOS varactor in the isolation port of each 90° hybrid coupler trims inter-path phase, while variable-gain amplifiers in the four quadrature branches provide ±2 dB tuning range to improve multi-level amplitude accuracy.

## IV. Fabrication and Experimental Results

The proposed transmitter is fabricated using GlobalFoundries' 90 nm SiGe BiCMOS process offering high speed HBTs with $f_T/f_{max}$ of 310/370 GHz. The die occupies 1.55 mm × 1.65 mm area and is wirebonded to a PCB for measurement as Fig. 4. The input LO is generated on chip by an ILO and its quadrupler path that consumes 49 mW DC power. The ILO is injected by Keysight 8257D analog signal generators with differential signals. Fig. 5(a) depicts the down-converted transient waveforms captured by a Keysight UXR0334B oscilloscope showing 180°, -22.5°, and 330° phase shifts created by analog varactor and digital cap bank, respectively. The transient waveform is compared to a reference signal as a trigger. The phase shifting versus frequency detuning under different injection conditions is depicted in Fig. 5(b), showing around 220 MHz ILO locking range and larger than ±360° phase shifting range. Fig. 5(c) displays the phase noise of the injected ILO and the signal generator, captured by FPL1026 spectrum analyzer. Due to the quadrupler and doubler path, the theoretical difference in phase noise between the transmitter output and the injection source is $20 \log_{10} 8 \approx 18$ dB, and the rms phase jitter is increased by a factor of 8. By integrating the measured phase noise over the desired bandwidth of 100 MHz and converting it to rms phase

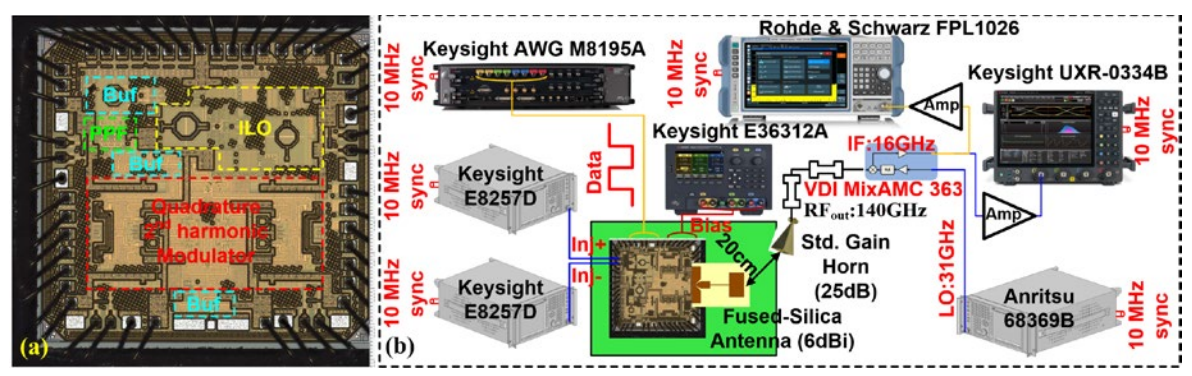

Fig. 4: (a) Chip micrograph in 90-nm SiGe BiCMOS. (b) Measurement setup.

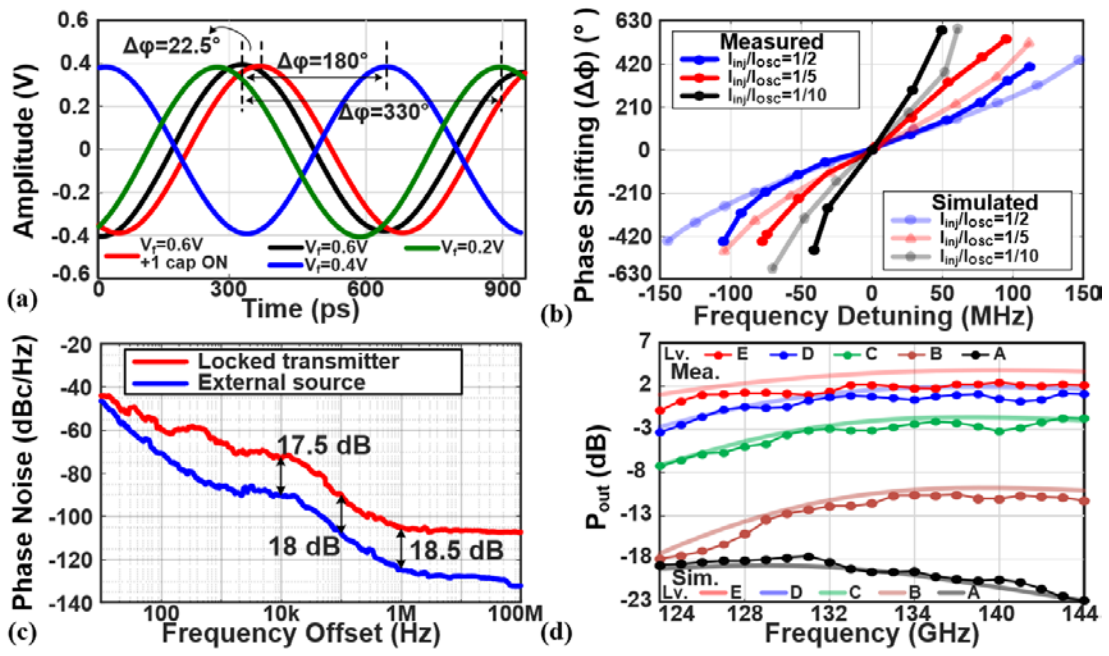

Fig. 5: (a) Measured transient waveforms demonstrating −22.5°, 180°, and 330° phase shifts via digital capacitor bank and analog varactor control; (b) Measured and simulated phase shift versus frequency detuning showing 220-MHz locking range and >±360° tuning; (c) Measured phase noise of locked transmitter output and external injection source through ×8 multiplication chain; (d) Measured output power versus frequency across five RC-OOK amplitude levels.

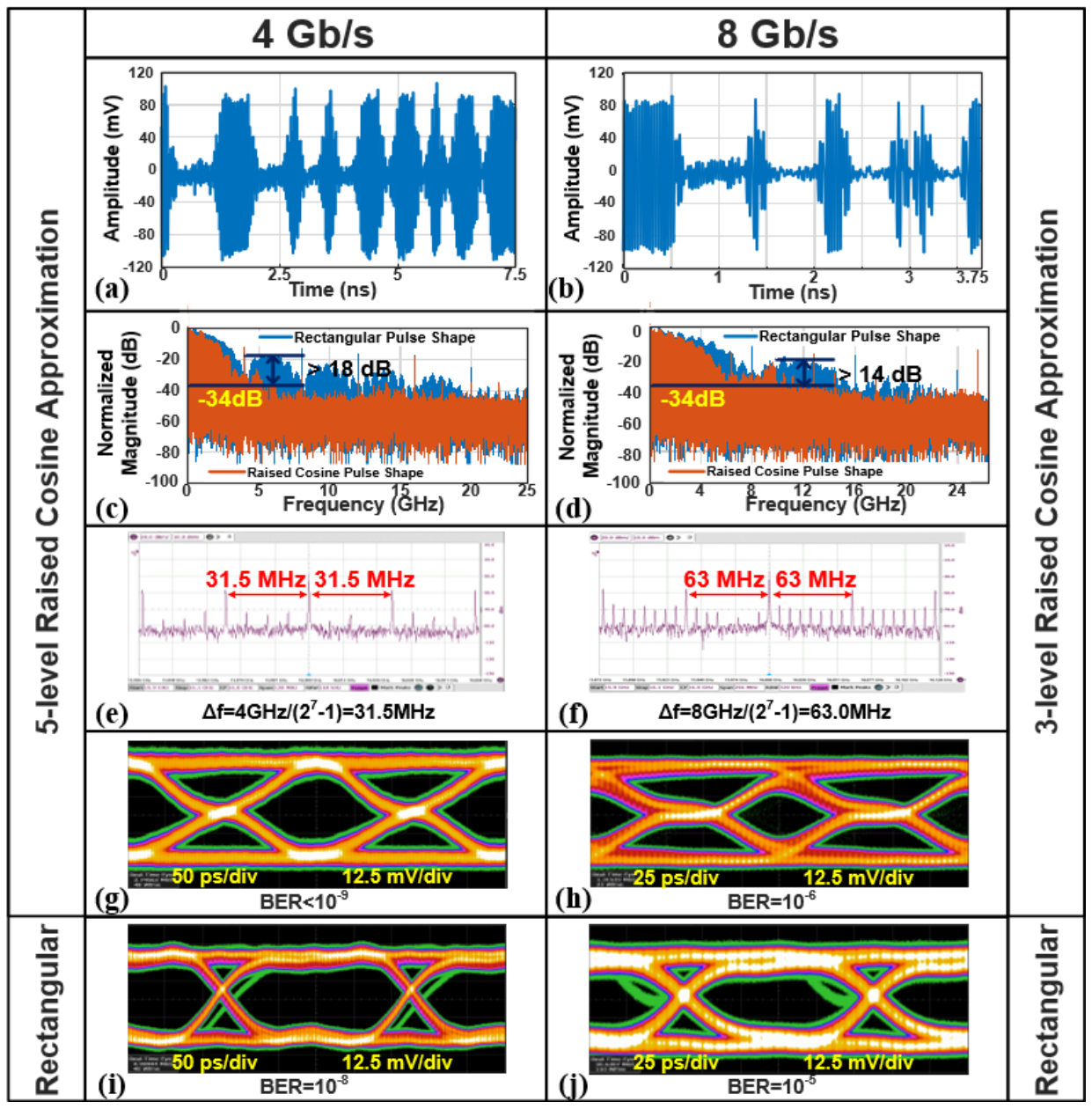

Fig. 6: Measured modulation results: (a), (b) down-converted transient waveforms for 4-Gb/s 5-level and 8-Gb/s 3-level operation, respectively; (c), (d) corresponding side-band comparisons; (e), (f) corresponding modulated spectra; (g), (h) corresponding eye diagrams over a 20-cm wireless link; (i), (j) eye diagrams of 4-Gb/s and 8-Gb/s rectangular NRZ.

error, the total rms phase uncertainty at the output can be calculated to be around 4.5 deg, corresponding to the minimum phase resolution. The maximum output power of the transmitter achieved is 2 dBm (Level E). Following staircase maximum power levels for Level D, C, B, A are 0 dBm, -3.5 dBm, -11 dBm and -21 dBm, forming an RC shape like waveform. The output power versus frequency curve, depicted in Fig. 5(d), indicates a full bandwidth across the ILO locking range. For

TABLE I: COMPARISON WITH SOA OOK MODULATOR AND PHASE SHIFTER

| | OOK Transmitter | | | Phase Shifter | | | This Work |
|---|---|---|---|---|---|---|---|
| Ref. | [12] | [5] | [4] | [13] | [14] | [15] | |
| Technology | 28 nm CMOS | 90 nm BiCMOS | 55 nm BiCMOS | 55 nm BiCMOS | 0.13 µm BiCMOS | 55 nm BiCMOS | **90 nm BiCMOS** |
| $f_C$ (GHz) | 135 | 140 | 225 | 150 | 115 | 150 | **140** |
| $P_{sat}$ (dBm) | 7 | 2 | 4.2[@] | 2 to 3.5[&] | -7.4[&] | -3.7[&] | **2** |
| Area ($mm^2$) | 0.975 | 0.32[!] | 0.89 | 0.2 | 2.108 | 0.05 | **2.56** |
| Osc. type/ Ph. Shifter type | VCO | External | Harmonic Oscillator | Vector-Sum | Vector-Sum | Vector-Sum | **Injection-locking** |
| LO Freq. (GHz) | 67.5 | 17.5 | 110 | N/A[2] | N/A[2] | N/A[2] | **17.5** |
| Multiplication | ×2 | ×8 | ×2 | N/A[2] | N/A[2] | N/A[2] | **×8** |
| Data Rate (Gb/s) | 20 (32)[%] | 12 | 20 | N/A[2] | N/A[2] | N/A[2] | **4[#]/8[*]** |
| $P_{DC}$ (mW) | 167 | 44[!] | 165.2 | 31 | 53 | 50 | **171** |
| TX Efficiency (pJ/bit) | 8.35 | 3.7[!] | 8.26 | N/A[2] | N/A[2] | N/A[2] | **21.4** |
| Side-lobe Rejection (dB) | N/A | N/A | N/A | N/A[2] | N/A[2] | N/A[2] | **34** |
| Distance (cm) | 6 | 20 | 1.5 (28)[$] | N/A[2] | N/A[2] | N/A[2] | **20** |
| Phase Range(°) | N/A[1] | N/A[1] | N/A[1] | 360 | 360 | 360 | **> ±360** |
| Phase Res. (°) | N/A[1] | N/A[1] | N/A[1] | 9 | 22.5 | Continuous | **22.5[<] /Continuous[>]** |
| RMS Gain Error (dB) | N/A[1] | N/A[1] | N/A[1] | 0.8 | 1.6 | 1.4 | **0.4** |
| RMS Phase Error (°) | N/A[1] | N/A[1] | N/A[1] | 5 | 5.5 | 7.5[§] | **4.5[?]** |

[%]: wire bonded [!]:Modulator only [@]:EIRP [$]:with lens [&]:OP1dB [§] :Measured across 32 states [1]: Not Phase Shifter
[2]: Not Tx [#]:5-level approx [*]:3-level approx [<]:Using cap bank [>]:Using analog varactor [?]:Phase uncertainty due to jitter

modulation measurements, a Keysight M8195A Arbitrary Waveform Generator (AWG) provides the PRBS baseband signal, Fig. 4(b). Fig. 6(a) and Fig. 6(b) show the measured down-converted 5-level and 3-level RC waveforms at 4 Gb/s and 8 Gb/s, respectively, with a 23-dB on/off ratio. On the receiver side, additional low-noise amplifier, envelope detection, and low-pass filtering were used to recover the data. The measured spectra in Fig. 6(c) and Fig. 6(d) show side-lobe suppression exceeding 34 dB for both the 4-Gb/s 5-level and 8-Gb/s 3-level RC-like envelope shaping cases. This corresponds to an additional suppression of 18 dB and 14 dB, respectively, compared to rectangular NRZ at the same data rates. Fig. 6(e) and Fig. 6(f) show the modulated spectra for 4 Gb/s 5-level RC approximation and 8 Gb/s for 3-level approximation, with Δf (symbol rate/($2^7$-1)) of 31.5 MHz and 63 MHz, respectively. Eye diagrams for 5-level RC approximation at 4 Gb/s, 3-level RC approximation at 8 Gb/s, rectangular NRZ for 4 Gb/s, and rectangular NRZ for 8 Gb/s, with 20 cm wireless link, are captured as depicted in Fig. 6(g), (h), (i), and (j). Compared with the rectangular NRZ, our proposed 5-level RC approximation shows better eye quality with less ISI and jitter, leading to better bit error rate of less than $10^{-6}$.

The comparison between the proposed work, prior OOK transmitters [12], [5], [4], and phase shifters [13]–[15], is summarized in Table I. The proposed transmitter combines 34-dB sidelobe suppression with a phase-tuning range larger than ±360°, making it attractive for scalable sub-THz array applications.

## V. CONCLUSION

A 140-GHz RC-OOK transmitter with direct RC envelope shaping and integrated ILO-based phase control has been demonstrated in 90-nm SiGe BiCMOS. By synthesizing multi-level RF amplitude states through a quadrature second-harmonic modulator, the architecture relaxes baseband pulse-shaping filters and high-speed DACs while achieving 34-dB sidelobe suppression at data rates up to 8 Gb/s with 2 dBm output power. The integrated D-band ILO phase shifter provides 22.5° digital resolution with continuous analog tuning beyond ±360°. These results make the proposed transmitter well suited to dense multi-channel short-range links, including data-center wireless interconnects and urban small-cell backhaul, as well as scalable phased-array where simultaneous spectral confinement and beam steering are required.

## ACKNOWLEDGMENT

The authors would like to thank GlobalFoundries for chip fabrication through HP University Shuttle Program, and Xiuhan Chen for re-simulating the ×4 LO path and assisting with layout verification.

## REFERENCES

[1] C. D'heer and P. Reynaert, "A Fully Integrated 135-GHz Direct-Digital 16-QAM Wireless and Dielectric Waveguide Link in 28-nm CMOS," in *IEEE Journal of Solid-State Circuits*, vol. 59, no. 3, pp. 889-907, March 2024, doi: 10.1109/JSSC.2023.3311673.

[2] H. Wang, *et al.*, "Analysis and Design of High-Order QAM Direct-Modulation Transmitter for High-Speed Point-to-Point mm-Wave Wireless Links," in *IEEE Journal of Solid-State Circuits*, vol. 54, no. 11, pp. 3161-3179, Nov. 2019, doi: 10.1109/JSSC.2019.2931610.

[3] C. -L. Cheng, S. Sangodoyin and A. Zajić, "THz Cluster-Based Modeling and Propagation Characterization in a Data Center Environment," in *IEEE Access*, vol. 8, pp. 56544-56558, 2020.

[4] B. Hadidian, *et al.*, "A 220-GHz Energy-Efficient High-Data-Rate Wireless ASK Transmitter Array," in *IEEE Journal of Solid-State Circuits*, vol. 57, no. 6, pp. 1623-1634, June 2022.

[5] H. Li, *et al.*, "Double HOOK: A 140 GHz 15 Gb/s Reconfigurable 3-Level ASK Modulator With Constant Input Impedance for High-Speed Connectivity," in *IEEE Microwave and Wireless Technology Letters*, vol. 35, no. 6, pp. 872-875, June 2025, doi: 10.1109/LMWT.2025.3557573.

[6] H. Li and N. Ebrahimi, "A 45 Gb/s D-band Hybrid Star-QAM-OOK Transmitter Using a Quad-Harmonic Modulator with Constant Impedance Balanced Architecture in 90nm SiGe BiCMOS," *2025 IEEE Radio Frequency Integrated Circuits Symposium (RFIC)*, San Francisco, CA, USA, 2025, pp. 167-170, doi: 10.1109/RFIC61188.2025.11082881.

[7] X. Chen, *et al.*, "CN-OOK: A Passive Common-Node OOK Modulator Using NLTL for Harmonic Generation at 120 GHz," *2025 IEEE BiCMOS and Compound Semiconductor Integrated Circuits and Technology Symposium (BCICTS)*, Phoenix, AZ, USA, 2025, pp. 1-4.

[8] N. Ebrahimi, *et al.*, "Direct Digital-to-Physical Synthesis: From Millimeter-Wave Transmitter to Qubit Control," in *IEEE Microwave Magazine*, vol. 27, no. 5, pp. 50-69, May 2026.

[9] M. Y. Yağan, *et al.*, "RF Chain-Free mmWave Transmission: Modeling and Experimental Verification," *2024 IEEE 35th International Symposium on Personal, Indoor and Mobile Radio Communications (PIMRC)*, Valencia, Spain, 2024, pp. 1-6.

[10] D. Widmann, *et al.*, "Mixed-Signal Integrated Circuit for Direct Raised-Cosine Filter Waveform Synthesis of Digital Signals up to 24 GS/s in 22 nm FD-SOI CMOS Technology," *2022 IEEE International Symposium on Circuits and Systems (ISCAS)*, Austin, TX, USA, 2022, pp. 3234-3238.

[11] N. Ebrahimi, *et al.*, "A 71–86-GHz Phased Array Transceiver Using Wideband Injection-Locked Oscillator Phase Shifters," in *IEEE Transactions on Microwave Theory and Techniques*, vol. 65, no. 2, pp. 346-361, Feb. 2017, doi: 10.1109/TMTT.2016.2647703.

[12] R. Zhang *et al.*, "A D-Band OOK Transmitter With 50-GHz Bandwidth Achieving 32-Gbps Data Rate in 28-nm CMOS," in *IEEE Journal of Solid-State Circuits*, vol. 59, no. 8, pp. 2347-2361, Aug. 2024.

[13] L. Piotto, G. De Filippi and A. Mazzanti, "33.3 A 125-to-170GHz Power-Efficient Phase Shifter in SiGe BiCMOS with Outphasing Gain and Phase Corrections," *2025 IEEE International Solid-State Circuits Conference (ISSCC)*, San Francisco, CA, USA, 2025, pp. 546-548.

[14] S. Afroz and K. -J. Koh, "A D -Band Two-Element Phased-Array Receiver Front End With Quadrature-Hybrid-Based Vector Modulator," in *IEEE Microwave and Wireless Components Letters*, vol. 28, no. 2, pp. 180-182, Feb. 2018, doi: 10.1109/LMWC.2018.2792150.

[15] D. d. Rio, *et al.*, "A Compact and High-Linearity 140–160 GHz Active Phase Shifter in 55 nm BiCMOS," in *IEEE Microwave and Wireless Components Letters*, vol. 31, no. 2, pp. 157-160, Feb. 2021.